\documentclass[prd,
,superscriptaddress,showpacs,nofootinbib,%
tightenlines]{revtex4}
\usepackage{epsfig}
\usepackage{amssymb,amsmath}
\usepackage{ulem}
\newcommand{\beq}{\begin{equation}}
\newcommand{\eeq}{\end{equation}}
\newcommand{\beqa}{\begin{eqnarray}}
\newcommand{\eeqa}{\end{eqnarray}}

\newcommand{\ben}{\begin{displaymath}}
\newcommand{\een}{\end{displaymath}}
\newcommand{\be}{\begin{equation}}
\newcommand{\ee}{\end{equation}}
\newcommand{\bea}{\begin{eqnarray}}
\newcommand{\eea}{\end{eqnarray}}
\begin{document}
\title{
Reply to "Comment on "How (not) to renormalize integral equations with
singular potentials in effective field theory"
}
\author{E.~Epelbaum}
\affiliation
{Ruhr-Universit\" at Bochum, Fakult\" at f\" ur Physik und Astronomie,  Institut f\" ur Theoretische Physik II, 44780 Bochum, Germany}
\author{A.~M.~Gasparyan}
\affiliation
{SSC RF ITEP, Bolshaya Cheremushkinskaya 25, 117218 Moscow, Russia}
\author{J.~Gegelia}
\affiliation{Ruhr-Universit\" at Bochum, Fakult\" at f\" ur Physik und Astronomie,  Institut f\" ur Theoretische Physik II, 44780 Bochum, Germany}
\affiliation{Tbilisi State  University,  0186 Tbilisi,
 Georgia}
\author{Ulf-G.~Mei\ss ner}
\affiliation{Helmholtz Institut f\"ur Strahlen- und Kernphysik and Bethe
   Center for Theoretical Physics, Universit\"at Bonn, D-53115 Bonn, Germany}
 \affiliation{Institute for Advanced Simulation, Institut f\"ur Kernphysik
   and J\"ulich Center for Hadron Physics, Forschungszentrum J\"ulich, D-52425 J\"ulich,
Germany}

\date{4 March, 2019}
\begin{abstract}
We offer a brief response to the criticism put forward by Pavon
Valderrama about our recent paper on ``How (not) to renormalize integral equations
with singular potentials in effective field theory''.

\end{abstract}

\maketitle
The criticism raised by Pavon Valderrama in
Ref.~\cite{Valderrama:2019yiv} concerns two issues summarized in the first
paragraph of that paper. Below, we address 
both issues and show that the arguments of Ref.~\cite{Valderrama:2019yiv} are of no
relevance for the conclusions reached in our paper~\cite{Epelbaum:2018zli}.


The first issue addressed by Pavon Valderrama concerns our statement about the
inconsistency of  taking the cutoff limit $\Lambda \to
\infty $ in non-perturbative expressions for the scattering
amplitude 
without having subtracted the relevant counterterms beforehand.  The
author does not point out any specific flaw in our
arguments  but simply  declares that  
our ``diagnosis is incorrect'' because ``redundant counterterms (RC)'',
apparently discussed in Ref.~\cite{Valderrama:2016koj}\footnote{We
  failed to find the definition of ``redundant counterterms" in
  Ref.~\cite{Valderrama:2016koj}.}, ``can be ignored in practice
when $\Lambda\to\infty$". He then claims that ``RCs are, however, 
regularly included in EFTs using the power divergence subtraction scheme
(PDS) regularization \cite{Kaplan:1998we}".   
Notice that Ref.~\cite{Kaplan:1998we} actually uses dimensional
regularization\footnote{Pavon Valderrama apparently confuses the
regularization and renormalization schemes.}  and therefore 
all loop integrals appearing in pionless EFT  are finite
in four space-time dimensions, i.e.~the scattering amplitude
considered in Ref.~\cite{Kaplan:1998we} has no residual cutoff
dependence. This example is thus of no relevance for our paper.
While these imprecise formulations and misleading statements make it
difficult to follow the arguments of  Ref.~\cite{Valderrama:2019yiv},
we have a feeling that the author has misinterpreted the statement in \cite{Epelbaum:2018zli} he is
objecting to.  We do by no means claim any inconsistency of removing
the cutoff in non-perturbative expressions for the scattering amplitude
provided one follows the steps: (i) calculate the amplitude regularized
with a cutoff $\Lambda$, (ii) subtract {\it all} ultraviolet
divergences in loop integrals emerging from iterations of the integral
equation and (iii) take the limit $\Lambda \to \infty$ afterwards. We do,
however, claim that performing (iii)  {\it without having carried out step
(ii)} generally leads to results which cannot be regarded as
renormalized and are incompatible with the principles of EFT, even if a
finite limit  $\Lambda \to \infty$ happens to exist for the 
amplitude. For pionless EFT, the algorithm specified above can indeed 
be easily implemented in the non-perturbative environment. In particular,
Eq.~(8) of our paper \cite{Epelbaum:2018zli} contains {\it all}
counterterms needed to remove the divergences from {\it all} terms in the
expansion of the amplitude in powers of $\hbar$ in Eq.~(6).  We are,
however, not aware of any calculations in spin-triplet nucleon-nucleon
channels based on a
non-perturbative treatment of the one-pion exchange potential, where
the step (ii) could be carried out (except for the approach proposed in
Ref.~\cite{Epelbaum:2012ua}).


Regarding the second issue, the author of
Ref.~\cite{Valderrama:2019yiv} has indeed succeeded to obtain a good
description of the toy-model phase shifts based on a perturbative
inclusion of  contact interactions for several values of the cutoff
parameter at the cost of fitting up to four adjustable
parameters.\footnote{Still, no evidence of the convergence with respect to the
coordinate-space cutoff $R_c$ and thus of the existence of the limit 
$R_c \to 0$ is provided for the phase shifts outside of the fitted
region.} However, as pointed out in our paper~\cite{Epelbaum:2018zli}, the large difference
between the full phase shifts and the leading-order (LO) ones suggests that the
results obtained from a perturbative inclusion of higher-order terms
are strongly dependent on the employed unitarization procedure, thus being
model-dependent. While repeating the
numerical analysis of Ref.~\cite{Valderrama:2019yiv} goes beyond the
scope of this comment, we can illustrate the origin of the problem
using the following simple considerations. 
We start with assuming that the perturbative expansion of
Ref.~\cite{Valderrama:2019yiv} is indeed convergent, i.e.~the full
scattering amplitude is well approximated by the  
first several terms in the perturbative expansion
\begin{equation}
T=T_{-1} \, \epsilon^{-1}+ T_1 \,\epsilon^1+ T_3\, \epsilon^3+ {\cal O}(\epsilon^5) ,
\label{tpert}
\end{equation}
where we introduced a parameter $\epsilon$ to keep track of orders in
small parameters (we set $\epsilon=1$ after the relative orders
are established). As $T_{-1}$ is a solution to the LO
integral equation, it is unitary by construction.  Thus,
$1/T-1/T_{-1}$ has to be
real. Expanding this expression in powers of $\epsilon$  
and demanding that each term is real, one can express the imaginary parts of
$T_1$ and $T_3$ in terms of their real parts and the LO
amplitude $T_{-1}$.   
Modulo higher-order corrections, the real parts of  $T_1$ and $T_3$
can then
be uniquely determined by demanding that the real and imaginary parts of
the toy-model amplitude are reproduced. According to Fig.~1 of
Ref.~\cite{Valderrama:2019yiv}, the phase shift  at e.g.~$k_{\rm
  cm}=0.22$ GeV  is accurately described at order $\nu=3$ for the
smallest considered cutoff $R_{\rm c}=0.3$~fm. Using the corresponding
numerical values of the LO and the full phase shifts $\delta_{\rm LO} =
-63.61^\circ$ and $\delta =-14.46^\circ$, 
we obtain two
solutions for the real parts of $T_1$ and $T_3$ corresponding to
the following   
expansion of the amplitude:
\begin{eqnarray}
T & = & (24.3\, -49.0\, i)\, \epsilon^{-1}+ (28.0\, +36.8\, i) \,\epsilon^1+ (-37.5\, +8.4\, i)\, \epsilon^3+ {\cal O}(\epsilon^5),\nonumber\\
T & = & (24.3\, -49.0 \, i) \, \epsilon^{-1}+ (-28.0-36.8\, i) \,\epsilon^1+ (18.4\, +82.0\, i)\, \epsilon^3+ {\cal O}(\epsilon^5) .
\label{twoampls}
\end{eqnarray}
Obviously, none of the expansions shows any sign of convergence.
Moreover, while both expressions do exactly reproduce the full, unitary amplitude $-(4\pi
/m )(1/T) = - 3.90 - i$ when
truncated at order $\epsilon^3$, none of the expressions for the
amplitude is approximately unitary when truncated at next-to-leading
order $\epsilon^1$. Specifically, we obtain $-(4\pi /m )(1/T_{\rm NLO}) =
-1.11 - 0.26 \,i$ for the first solution, while $-(4\pi /m )(1/T_{\rm NLO}) = 0.03 - 0.71 \,i$
for the second one. 
Using different unitarization prescriptions one can, in fact, obtain a
broad range of phase shifts at NLO including the ones shown in Fig.~1
of Ref.~\cite{Valderrama:2019yiv}.  
It is important to stress that the aim of an EFT is, however,  not to describe the data at any price
but rather to provide a systematic approach with controlled accuracy
and reliable error estimations.  

\section*{Acknowledgments}

This work was supported in part by the Georgian Shota Rustaveli National
Science Foundation (Grant No. FR17-354),  by DFG (SFB/TR 110, ``Symmetries and the Emergence of Structure in
QCD'') and the BMBF  (Grant No. 05P18PCFP1). Further support was provided by the Chinese
Academy of Sciences (CAS) President's International
Fellowship Initiative (PIFI) (grant no. 2018DM0034) and by VolkswagenStiftung
(grant no. 93562).


\begin{thebibliography}{1}


\bibitem{Valderrama:2019yiv} 
  M.~P.~Valderrama,
  arXiv:1901.10398 [nucl-th].

\bibitem{Epelbaum:2018zli} 
  E.~Epelbaum, A.~M.~Gasparyan, J.~Gegelia and U.-G.~Mei\ss ner,
  Eur.\ Phys.\ J.\ A {\bf 54}, no. 11, 186 (2018).

\bibitem{Valderrama:2016koj} 
  M.~P.~Valderrama,
  Int.\ J.\ Mod.\ Phys.\ E {\bf 25}, no. 05, 1641007 (2016).

\bibitem{Kaplan:1998we} 
  D.~B.~Kaplan, M.~J.~Savage and M.~B.~Wise,
  Nucl.\ Phys.\ B {\bf 534}, 329 (1998).

  \bibitem{Epelbaum:2012ua} 
  E.~Epelbaum and J.~Gegelia,
  Phys.\ Lett.\ B {\bf 716}, 338 (2012).

  
\end{thebibliography}
\end{document}